\documentclass[aps, prd, twocolumn, twoside, 10pt, nofootinbib,
flushbottom]{revtex4-1}
\usepackage{amsmath,amssymb}
\usepackage{graphicx}
\usepackage{txfonts}
\usepackage{bbm}
\usepackage{slashed}

\newcommand{\Nf}{{N_\text f}}                             
\newcommand{\Nfc}{N_\text {f}^\text{cr}}                  
\newcommand{\iu}{\mathrm i}                               
\newcommand{\bpsi}{\bar\psi}
\newcommand{\euler}[1]{\ee^{#1}}                          
\newcommand{\Eqref}[1]{Eq.~\eqref{#1}}
\DeclareMathOperator{\ee}{e}                              
\DeclareMathOperator{\Tr}{Tr}                             
\DeclareMathOperator{\Intdq}{\int^\Lambda_0 d \mathit{|q|}}
\DeclareMathOperator{\Intdx}{\int d^3 \mathit{x}}

\begin{document}
\title{UV fixed-point structure of the three-dimensional
Thirring model}
\date{\today}
\author{Holger Gies}
\email{holger.gies@uni-jena.de}
\author{Lukas Janssen}
\email{lukas.janssen@uni-jena.de}
\affiliation{Theoretisch-Physikalisches Institut,
Friedrich-Schiller-Universit{\"a}t Jena, Max-Wien-Platz 1, 07743
Jena, Germany}

\begin{abstract}
  We investigate the UV fixed-point structure of the three-dimensional
  Thirring model by means of the functional renormalization group (RG). We
  classify all possible 4-fermi interactions compatible with the present
  chiral and discrete symmetries and calculate the purely fermionic RG flow
  using a full basis of fermionic four-point functions in the point-like
  limit. Our results show that the UV complete (asymptotically safe) Thirring
  model lies in a two-dimensional coupling plane which reduces to the
  conventional Thirring coupling only in the strict large-$\Nf$ limit. In
  addition to the Thirring universality class which is characterized by one
  relevant direction (also at finite $\Nf$), two further interacting fixed
  points occur which may define new universality classes of second-order phase
  transitions also involving parity-broken phases.  The $\Nf$-dependence of
  the Thirring fixed point sheds further light on the existence of an
  $\Nf$-controlled quantum phase transition above which chiral symmetry
  remains unbroken for arbitrary large coupling, even though a definite answer
  will require a direct computation of competing orders.
\end{abstract}
\maketitle

\section{Introduction}
Three-dimensional relativistic fermionic systems have extensively been
investigated in the literature in a variety of scenarios.  On the one hand,
they are per se interesting field theories with unconventional features, on
the other hand they allow for fascinating applications to condensed-matter
systems. In particular, three-dimensional quantum electrodynamics (QED$_3$) or
the Thirring model \cite{Thirring:1958in} are actively discussed, e.g., as
effective theories describing different regions of the phase diagram of
high-$T_\text c$ cuprate superconductors \cite{Franz:2001zz, Herbut:2002wd,
  Mavromatos:2003ss} and, recently, the electronic properties of graphene
\cite{Semenoff:1984dq, Herbut:2006cs, Herbut:2009qb, Drut:2008xx,
  Gusynin:2007ix}. Especially graphene since its first
synthesis in 2004 \cite{Novoselov:2004xx} is being lively discussed in the
rapidly growing literature on this subject, also because it offers the appealing
opportunity of a comparatively simple experimental realization of some up to now
unobserved quantum relativistic phenomena, such as the Klein paradox
\cite{Klein:1929zz} or Zitterbewegung \cite{Schroedinger:1930xx}; for reviews,
see \cite{Geim:2007xx, CastroNeto:2007xx}. More exotically, some features of
3$d$ relativistic fermion systems can serve as toy models, e.g., for the
standard model of particle physics \cite{Gies:2009da}, or a possible candidate
for a ``Theory of Everything'' \cite{Alexandre:2009iu}.

However, QED$_3$ \cite{Pisarski:1984dj, Cornwall:1980zw, Maris:1996zg,
  Kubota:2001kk, Hands:2002dv, Gusynin:2003ww, Fischer:2004nq} and the
three-dimensional Thir\-ring model \cite{Parisi:1975im, Hands:1994kb,
  Gomes:1990ed, Hong:1993qk, Itoh:1994cr, Kondo:1995jn, Kim:1996xza,
  Christofi:2007ye} are likewise intrinsically interesting, in particular
because the ground states in these theories are expected to show a sensitivity
to the number of fermion flavors $\Nf$. Several approximate solutions of the
Dyson-Schwinger equations (DSE) predict that chiral symmetry breaking ($\chi$SB)
is prohibited once $\Nf$ is larger than a critical value $\Nfc$
\cite{Gomes:1990ed, Itoh:1994cr, Kondo:1995jn}. A similar quantum critical
phenomenon has also been identified in many-flavor QCD$_4$
\cite{Miransky:1996pd, Gies:2005as}, being currently under intense investigation
also because of its potential relevance for technicolor scenarios
\cite{Ryttov:2007cx} and its implications for the thermal phase boundary
\cite{Braun:2009ns}. 

The search for the quantum critical point in the Thirring model has so far
been rather challenging: different DSE approaches, e.g., have yielded values
between $\Nfc \simeq 3.24$ \cite{Gomes:1990ed} and $\Nfc = \infty$
\cite{Hong:1993qk}. Recent extensive lattice simulations now point to $\Nfc
\simeq 6.6$ \cite{Christofi:2007ye}. For physical graphene and cuprates, the
number of flavors is $\Nf=2$, such that the true value of $\Nfc$ may be of 
profound importance for physical effects corresponding to chiral symmetry
breaking in the effective theories. In fact, a dynamically induced mass gap in
the band structure of graphene, corresponding to a semimetal-Mott insulator
quantum phase transition, could provide extraordinary applications for
graphene-based electronics, offering a possible new candidate material to take
over from Si-based technology \cite{Geim:2007xx}.

In this work, we take a more fundamental viewpoint in order to investigate the
amount of universality that can be attached to a possible value for
$\Nfc$. Naively, a universal answer for $\Nfc$ may not be apparent as 3$d$
fermion models are perturbatively nonrenormalizable. Nevertheless,
renormalizability of the 3$d$ Thirring model to any order in a large-$\Nf$
expansion has been shown in \cite{Parisi:1975im, Hands:1994kb, Gomes:1990ed,
  Hong:1993qk}, with a diagrammar being very similar to QED$_3$. This has been
taken as an indication that the Thirring model can indeed by defined
nonperturbatively in 3$d$, providing the same amount of universality as any
perturbatively renormalizable theory. 

Universality and, more profoundly, UV completeness can, in fact, be analyzed
within Weinberg's scenario of asymptotic safety \cite{Weinberg:1980gg,
  Niedermaier:2006wt}, in which a UV complete infinite-cutoff limit can be taken
at a potentially non-Gau\ss ian fixed point. The resulting theories are
predictive and exhibit universality, if the number of RG relevant directions at
the fixed point are finite. In this work, we re-examine renormalizability of the
3$d$ Thirring model from this viewpoint by means of the functional
renormalization group formulated in terms of the Wetterich equation
\cite{Wetterich:1992yh}. Using a purely fermionic RG as a first step, the UV
fixed-point structure can indeed be mapped out, on the one hand confirming the
large-$\Nf$ results, but also revealing interesting deviations from the
large-$\Nf$ asymptotics.  For instance, we can identify a non-Gau\ss ian UV
fixed point that defines a ``Thirring universality class'' which, however,
corresponds to a pure Thirring coupling only in the strict large-$\Nf$ limit. In
addition, we find further fixed points which may be associated with phase
transitions (and corresponding new universality classes), e.g., towards a parity
broken phase.

This paper is organized as follows. In Sec.~\ref{sec:symmetries}, we analyze
the discrete and continuous symmetries of the classical theory. We classify
the different possible mass terms as well as all possible fermionic
interaction channels in the point-like four-fermion limit. With the aid of the
functional RG, we investigate the purely fermionic RG flow in
Sec.~\ref{sec:rg-flow}, followed by an analysis of the fixed-point structure
of the given class of models for varying number of fermion flavors in
Sec.~\ref{sec:fixed-points}. In Sec.~\ref{sec:discussion}, we classify the
resulting RG trajectories to the long-range physics, pointing to the existence
of various phases with dynamical mass generation, and obtain a first glance at
a possible mechanism for the formation of a critical flavor number, above
which chiral symmetry breaking disappears. We give our conclusions in
Sec.~\ref{sec:conclusions}.

\section{Symmetries of the Thirring model} \label{sec:symmetries}
The Lagrangian for the massless Thirring model in three
Euclidean space-time dimensions is
\begin{equation}\label{eq:lagrangian}
\mathcal L = \bar\psi^a \iu\slashed{\partial}\psi^a 
+ \frac{\bar{g}}{2\Nf}\left(\bar\psi^a \gamma_\mu \psi^a\right)^2,
\end{equation}
satisfying Osterwalder-Schrader positivity.\footnote{For Osterwalder-Schrader
  positivity \cite{Osterwalder:1973dx}, we require invariance of the action
  under (generalized) complex
  conjugation defined by $\psi^\dagger \coloneqq \iu \bpsi \gamma_3$ together
  with reflection of the (euclidean) time coordinate (which we choose to
  be~$x_3$). For a detailed discussion of our chiral conventions, see e.g.,
  Ref.~\cite{Wetterich:2010ni}.}  The index $a$ runs over $\Nf$ distinct
fermion flavors. In three dimensions, we could use $2\times 2$ matrices (e.g.,
the Pauli matrices) as an irreducible representation of the Euclidean Dirac
algebra
\begin{equation}
\left\{\gamma_\mu,\gamma_\nu\right\} = 2 \delta_{\mu\nu},
\end{equation}
but this representation does not permit a chiral symmetry.
We therefore work exclusively with a $4\times 4$ \emph{reducible}
representation of the Dirac algebra; an explicit representation is given by
\begin{equation}
\gamma_\mu =
\begin{pmatrix}
0 & -\iu \sigma_\mu \\
\iu \sigma_\mu & 0
\end{pmatrix}, \qquad \mu=1,2,3,
\end{equation}
with $\{\sigma_1,\sigma_2,\sigma_3\}$ being the $2\times 2$ Pauli matrices.
Thus $\psi^a$ represents a \emph{four-component} Dirac spinor. Now there are 
\emph{two} further $4\times 4$ matrices which anticommute with all $\gamma_\mu$
as well as with each other,
\begin{equation}
\gamma_4 =
\begin{pmatrix}
0 & \mathbbm 1 \\
\mathbbm 1 & 0
\end{pmatrix} \quad \text{and} \quad \gamma_5 =
\gamma_1\gamma_2\gamma_3\gamma_4 =
\begin{pmatrix}
\mathbbm 1 & 0 \\
0 & -\mathbbm 1
\end{pmatrix}.
\end{equation}
The massless Lagrangian \eqref{eq:lagrangian} then is invariant under
the axial transformations
\begin{align}
\mathrm U_{\gamma_4}(1):\qquad
\psi &\mapsto \euler{\iu \alpha \gamma_4} \psi, &
\bar\psi &\mapsto \bar\psi \euler{\iu \alpha \gamma_4}, \\
\mathrm U_{\gamma_5}(1):\qquad
\psi &\mapsto \euler{\iu \beta \gamma_5} \psi, &
\bar\psi &\mapsto \bar\psi \euler{\iu \beta \gamma_5},
\end{align}
as well as the vector transformations
\begin{align}
\mathrm U_{\mathbbm 1}(1):\qquad
\psi &\mapsto \euler{\iu \vartheta} \psi, &
\bar\psi &\mapsto \bar\psi \euler{-\iu \vartheta}, \\
\mathrm U_{\gamma_{45}}(1):\qquad
\psi &\mapsto \euler{\iu \varphi \gamma_{45}} \psi, &
\bar\psi &\mapsto \bar\psi \euler{-\iu \varphi \gamma_{45}},
\end{align}
with $\gamma_{45} \coloneqq \iu\gamma_4\gamma_5$.
For each flavor $a=1,\dots,\Nf$, the theory thus has a global $\mathrm U(2)$
symmetry with the Hermitian generators $\tau_j$, $j=1,\dots,4$,
\begin{equation}\label{eq:U2-generators}
\tau_j = \mathbbm 1, \gamma_4, \gamma_5, \gamma_{45}.
\end{equation}
This symmetry transformations together with flavor rotations form a larger
symmetry of the classical massless Lagrangian, corresponding to the group
$\mathrm U(2\Nf)$ with the $(2\Nf)^2$ generators
\begin{equation}\label{eq:U2N-generators}
\lambda_i \otimes \tau_j, \qquad i=1,\dots,\Nf^2,\quad j=1,\dots,4.
\end{equation}
Here, $\{\lambda_1,\dots,\lambda_{\Nf^2-1}\}$ are the generalized $\Nf\times
\Nf$ Gell-Mann matrices, and $\lambda_{\Nf^2} \coloneqq \mathbbm 1_{\Nf}$ is
the identity. In other words, combining the $\Nf$ four-component spinors
$\psi^a = (\psi^a_\mathrm L,\psi^a_\mathrm R)^\mathrm T$ (each consisting of 2
two-component Weyl spinors $\psi_\mathrm L, \psi_\mathrm R$) into one
collective $4\Nf$-component spinor
$$\Psi\coloneqq\left(\psi^1_\mathrm L,\psi^1_\mathrm R,\dots,\psi^\Nf_\mathrm
L,\psi^\Nf_\mathrm R\right)^\mathrm T$$
(consisting of $2\Nf$ two-component Weyl spinors), the theory is invariant
under
\begin{equation}
\Psi \mapsto \mathcal U \Psi, \qquad \bar\Psi \mapsto \bar\Psi \mathcal
U^\dagger, \qquad \mathcal U\in \mathrm U(2\Nf),
\end{equation}
where the entries of the unitary $(2\Nf)\times (2\Nf)$ matrix $\mathcal U$ are
complex numbers times the $2\times 2$ identity matrix $\mathbbm 1_{2}$.

Because of the reducible representation of the Dirac algebra, discrete
transformations can be implemented in various ways \cite{Gomes:1990ed}. Let us
define
\begin{align*}
\mathcal C&:\ \psi \mapsto (\bar\psi C_\xi)^\mathrm T,\ \bar\psi \mapsto -
  (C_\xi^\dagger \psi)^\mathrm T &&\text{(charge conjugation)}, \\
\mathcal P&:\ \psi \mapsto P_\zeta \psi,\ \bar\psi \mapsto \bar\psi
  P_\zeta^\dagger &&\text{(parity inversion)}, \\
\mathcal T&:\ \psi \mapsto T_\eta\psi,\ \bar\psi \mapsto \bar\psi T_\eta^\dagger
  &&\text{(time reversal)},
\end{align*}
where $\mathcal C$ and $\mathcal P$ are unitary and $\mathcal T$ is
antiunitary. The arguments of the transformed fields are $\tilde x \coloneqq
(-x_1,x_2,x_3)$ in the case of parity, and $\hat x \coloneqq (x_1,x_2,-x_3)$ in
the case of time reversal. $C_\xi$, $P_\zeta$, and $T_\eta$ are unitary $4\times
4$ matrices given by
\begin{align}\label{eq:C-matrix}
C_\xi &= \frac{1}{2} \left[(1+\xi)\gamma_2\gamma_4 + \iu
(1-\xi)\gamma_2\gamma_5 \right], \\ \label{eq:P-matrix}
P_\zeta & = \frac{1}{2} \left[(1+\zeta)\gamma_1\gamma_4 + \iu
(1-\zeta)\gamma_1\gamma_5 \right], \\ \label{eq:T-matrix}
T_\eta & = \frac{1}{2} \left[(1+\eta)\gamma_1 + \iu
(1-\eta)\gamma_2\gamma_3\right],
\end{align}
depending on the pure phases $\xi$, $\zeta$, $\eta$ with
$|\xi|=|\zeta|=|\eta|=1$. Recall that $\gamma_1$ and $\gamma_3$ are
antisymmetric and purely imaginary, whereas $\gamma_2$, $\gamma_4$, and
$\gamma_5$ are symmetric and real. We thus see that on the classical level our
theory is invariant under any of the discrete transformations $\mathcal C$,
$\mathcal P$, and $\mathcal T$ individually, irrespective of the values of the
phases $\xi$, $\zeta$, and $\eta$.

Let us now consider building blocks of the effective action, starting at the
two-fermion level.  There are, in fact, four possible mass terms
$\bpsi\tau_j\psi$ with $\tau_j$ given in Eq.~\eqref{eq:U2-generators} and
diagonal flavor structure. (We shall suppress the flavor index $a$ as long as
it is not needed.) However, the term $\bpsi(\iu m + m' \gamma_4) \psi$
transforms under $\psi\mapsto\euler{\iu\alpha\gamma_4} \psi$ into a parity
even mass term $\propto \bpsi \psi$ if $\alpha$ is chosen to satisfy $2\alpha
= \arctan(m'/m)$. The analogous statement holds for the mass term
involving $\gamma_5$. More generally, any mass term can be transformed by a
$\mathrm U(2\Nf)$ rotation into\footnote{Note that due to our chiral
  conventions a nonzero expectation value $\langle\bpsi\psi\rangle$ or
  $\langle \bpsi\gamma_{45}\psi\rangle$ is purely imaginary
  \cite{Wetterich:2010ni}.}
\begin{equation}
\iu \bpsi \left(m + \tilde m \gamma_{45} \right) \psi.
\end{equation}
A dynamically generated mass $m \neq 0$ spontaneously breaks the $\mathrm
U(2\Nf)$ symmetry down to a residual $\mathrm{U}_{\mathbbm 1+\gamma_{45}}(\Nf)
\otimes \mathrm{U}_{\mathbbm 1-\gamma_{45}}(\Nf) \varsubsetneq \mathrm
U(2\Nf)$ generated by $\lambda_b\otimes (\mathbbm 1 \pm \gamma_{45})$,
$b=1,\dots,\Nf^2$ (c.f. Eq.~\eqref{eq:U2N-generators}), but leaves the discrete
space-time symmetries $\mathcal C$, $\mathcal P$, and $\mathcal T$
intact, in agreement with the analogous discussion in the context of 
QED$_3$~\cite{Pisarski:1984dj}. A nonvanishing mass $\tilde m$ in contrast does
\emph{not} break the $\mathrm U(2\Nf)$ symmetry, since $\gamma_{45}$
anticommutes with $\gamma_4$ and $\gamma_5$. However, as can be read off from
Eqs.~(\ref{eq:C-matrix}--\ref{eq:T-matrix}), such a mass term is odd under
parity inversion since $\gamma_{45}$ anticommutes with $P_\zeta$. Because of
$\{\gamma_{45},C_\xi\}=0$ and $(\gamma_{45})^\mathrm T = -\gamma_{45}$ it is
even under charge conjugation; since $[\gamma_{45},T_\eta]=0$ and
$(\iu\gamma_{45})^* = \iu \gamma_{45}$ it is also even under time reversal.

A complete basis of the $4\times 4$ Dirac algebra is given by the
$16$ matrices
\begin{equation} \label{eq:basis_dirac}
\left\{\gamma^A\right\}_{A=1}^{16} = \left\{\mathbbm 1, \gamma_\mu,
\sigma_{\mu\nu}/\sqrt{2},
\iu\gamma_\mu\gamma_4,
\iu\gamma_\mu\gamma_5, \gamma_4, \gamma_5, \gamma_{45}\right\},
\end{equation}
where we have introduced the generators of the Lorentz transformation of the
four-component Dirac spinors $\sigma_{\mu\nu}\coloneqq \frac{\iu}{2}
[\gamma_\mu, \gamma_\nu]$. In \Eqref{eq:basis_dirac}, we only count those
matrices $\sigma_{\mu\nu}$ with $\mu<\nu$. A bilinear $\bpsi \gamma^A \psi$ is
invariant under $\mathrm U(2\Nf)$ transformations if and only if $\gamma^A$
anticommutes with the generators of $\mathrm U_{\gamma_{4}}(1)$ and $\mathrm
U_{\gamma_{5}}(1)$ while it commutes with the generators of $\mathrm
U_{\mathbbm 1}(1)$ and $\mathrm U_{\gamma_{45}}(1)$. Obviously, this is only
the case for $\bpsi\gamma_\mu\psi$ and $\bpsi\gamma_{45}\psi$. Imposing an
invariance under $\mathrm U(2\Nf)$ as well as $\mathcal C$, $\mathcal P$, and
$\mathcal T$, there is no bilinear to zeroth derivative order. In
particular, no mass term is permitted. To first order, only the standard kinetic
term
\begin{equation}
\mathcal{L}_{\text{kin}}= \iu \bpsi\slashed{\partial}\psi
\end{equation}
can appear. Consequently, on the level of four-fermion interactions, the
Thirring interaction is not the only fermionic $4$-point function in the
point-like limit (i.e., with momentum independent couplings) which is invariant
under the present $\mathrm U(2\Nf)$ flavor symmetry and the discrete space-time
symmetries. In fact, the possible interactions are
\begin{equation} \label{eq:invariant_4-fermi}
S_\mu^2 \coloneqq \left(\bpsi^a\gamma_\mu\psi^a\right)^2 \quad \text{and} \quad
S^2  \coloneqq \left(\bpsi^a\gamma_{45}\psi^a\right)^2,
\end{equation}
as well as the two interaction terms with non-singlet flavor
structure
\begin{equation} \label{eq:invariant_4-fermi_fierz-A}
V^2 \coloneqq \left(\bpsi^a\psi^b\right)^2 -
\left(\bpsi^a\gamma_4\psi^b\right)^2 -
\left(\bpsi^a\gamma_5\psi^b\right)^2 + \left(\bpsi^a\gamma_{45}\psi^b\right)^2,
\end{equation}
\begin{equation}
\label{eq:invariant_4-fermi_fierz-B}
V_\mu^2 \coloneqq \left(\bpsi^a\gamma_\mu\psi^b\right)^2 +
\left(\bpsi^a\frac{\sigma_{\mu\nu}}{\sqrt{2}}\psi^b\right)^2 
- \left(\bpsi^a\iu \gamma_\mu\gamma_4\psi^b\right)^2
- \left(\bpsi^a\iu \gamma_\mu\gamma_5\psi^b\right)^2,
\end{equation}
where we define $(\bpsi^a\psi^b)^2 \equiv \bpsi^a\psi^b\bpsi^b\psi^a$, etc.
However, the terms in
Eqs.~\eqref{eq:invariant_4-fermi_fierz-A}, \eqref{eq:invariant_4-fermi_fierz-B}
are not independent of those in \Eqref{eq:invariant_4-fermi} but can be mapped
onto each other by means of Fierz transformations. We find that the
scalar/pseudoscalar interaction in \Eqref{eq:invariant_4-fermi_fierz-A} is equal
to $V^2 = -S_\mu^2 - S^2$ and the axial/vector-type interaction
in~\eqref{eq:invariant_4-fermi_fierz-B} obeys $V_\mu^2 =
S_\mu^2 - 3 S^2$; see Appendix~\ref{app:fierz}. This generalizes the discussion
of \cite{Herbut:2009qb} to larger flavor number~$\Nf$.

To summarize: in addition to the Thirring coupling $S_\mu^2$, a second
point-like linearly independent four-fermi coupling $S^2$ satisfies the
symmetries of the Thirring model. In an RG analysis, it has to be included on
the same fundamental level as the Thirring interaction.

\section{Fermionic RG flow} \label{sec:rg-flow}
All physical information of a quantum field theory is stored in correlation
functions which in turn can be extracted from a generating functional. By a
Legendre transform of the latter one obtains the \emph{effective action}
$\Gamma$, which governs the dynamics of the field expectation value, taking
the effects of all quantum fluctuations into account.
 In other words, a given theory is solved, once $\Gamma$ is computed. Instead of
integrating out all
fluctuations at once, we can implement Wilson's idea of integrating
out modes momentum shell by momentum shell, leading us to the scale dependent
\emph{effective average action} $\Gamma_k$, with a momentum-shell parameter
$k$. $\Gamma_k$ corresponds to the bare action $S=\Intdx \mathcal L$
for $k$ approaching the UV cut-off $\Lambda$, while the full
quantum action $\Gamma$ is approached for $k\rightarrow 0$. The scale dependence
of $\Gamma_k$ (as a functional of only fermionic degrees of freedom in our case)
is governed by the Wetterich equation~\cite{Wetterich:1992yh}
\begin{equation} \label{eq:wetterich-eq}
\partial_t \Gamma_k[\bpsi,\psi] = - \frac{1}{2} \Tr \left(\frac{\partial_t
R_k}{\Gamma_k^{(2)}[\bpsi,\psi] + R_k}\right), \qquad \partial_t \equiv
k \frac{\partial}{\partial k},
\end{equation}
where the trace is meant to be taken over all internal degrees of freedom
(flavor, spinor, momentum). Here, $\Gamma_k^{(2)}[\bpsi,\psi]$ is the second
functional derivative with respect to $\bpsi$ and $\psi$, and $R_k$ denotes a
momentum-dependent regulator function, ensuring that IR modes below the
momentum scale $k$ are suppressed. The minus sign on the right hand side of
Eq.~\eqref{eq:wetterich-eq} is due to the Grassmann nature of $\bpsi$ and
$\psi$. For reviews, see e.g.\ \cite{Berges:2000ew, Kopietz:2010xx}.  
With the Wetterich equation being an exact equation, consistent approximation
schemes can be devised that allow for a systematic nonperturbative investigation
of the given model. In this work, we use a simple derivative expansion of the
effective action in terms of purely fermionic degrees of freedom with point-like
interactions,
\begin{multline} \label{eq:truncation}
\Gamma_k[\bpsi,\psi] = \Intdx \left\{ 
Z_k \bpsi^a \iu \slashed{\partial} \psi^a + 
\frac{\tilde{\bar{g}}_k}{2\Nf} \left(\bpsi^a\gamma_{45}\psi^a\right)^2 \right.\\
\left. + \frac{\bar{g}_k}{2\Nf} \left(\bpsi^a\gamma_\mu\psi^a\right)^2
\right\}.
\end{multline}
In addition to the interaction terms discussed above, we have included a wave
function renormalization $Z_k$. All parameters in the effective average action
are understood to be scale dependent which is indicated by the momentum-scale
index $k$. This truncation corresponds to a next-to-leading order derivative
expansion, which can consistently be extended to higher orders and thus
defines a systematic nonperturbative approximation scheme. As
discussed in the previous section, the truncation \eqref{eq:truncation}
represents a full basis of fermionic 4-point functions in the point-like
limit, which are compatible with the present chiral and discrete symmetries.
Such a point-like truncation can be a reasonable approximation in the chirally
symmetric regime, as has been quantitatively confirmed for the zero-temperature
chiral phase transition in many flavor QCD~\cite{Gies:2005as}.

Inserting Eq.~\eqref{eq:truncation} into Eq.~\eqref{eq:wetterich-eq}, we
obtain the flow equations (i.e., $\beta$ functions) for the 4-fermi couplings
$\bar{g}_k$ and $\tilde{\bar{g}}_k$ and the wave function renormalization $Z_k$
via suitable projections onto the associated operators.  For the explicit
computations, we refer the reader to Appendix~\ref{app:flow_eqs}. In terms of
renormalized fields
\begin{equation}
\psi \mapsto Z_k^{-1/2} \psi_k, \qquad \bpsi \mapsto Z_k^{-1/2}\bpsi_k,
\end{equation}
and dimensionless renormalized couplings
\begin{equation} \label{eq:dimensionless_couplings}
g = Z_k^2 k^{-1} \bar{g}_k, \qquad \tilde{g} = Z_k^2 k^{-1} \tilde{\bar{g}}_k,
\end{equation}
we obtain the beta functions as
\begin{align} \label{eq:beta-function_tilde-g}
\partial_t \tilde g & =\tilde g - \frac{4\ell_1^{(\mathrm F)}}{\pi^2}
\left[\frac{2\Nf-1}{2\Nf} \tilde g^2 - \frac{3}{2\Nf}
\tilde g g - \frac{1}{\Nf} g^2\right], \\
\label{eq:beta-function_g}
\partial_t g & = g + \frac{4\ell_1^{(\mathrm F)}}{\pi^2} 
\left[ \frac{1}{2\Nf} \tilde g g + \frac{2\Nf + 1}{6\Nf} g^2 \right].
\end{align}
Within the present truncation of point-like interactions, the anomalous
dimension remains $\eta_k \coloneqq -\partial_t \ln Z_k \equiv 0$. These flows
involve the threshold function $\ell_1^{(\mathrm F)}$ which encodes the
details of the regularization scheme as specified by the dimensionless
regulator shape function $r(q^2/k^2)$ defined by $R_k(q)= Z_k \slashed{q}
r(q^2/k^2)$,
\begin{equation} \label{eq:threshold-constant}
\ell_1^{(\mathrm F)} \coloneqq - \frac{\partial}{\partial k} \Intdq 
\frac{1}{\left[1+r\left(q^2/k^2\right)\right]^2}.
\end{equation}
For a given regulator function, this integral in the present truncation boils
down to a simple number. For instance, for the sharp cut-off
\begin{equation}
r^{\mathrm{sc}}(q^2/k^2) =
\begin{cases}
\infty & \text{for } q^2<k^2, \\
0 & \text{for } q^2>k^2,
\end{cases}
\end{equation}
we obtain $\ell_1^{(\mathrm F)}=1$. For a linear cut-off which satisfies a
regulator optimization criterion~\cite{Litim:2001up},
\begin{equation}
r^{\mathrm{opt}}(q^2/k^2) = \left(\sqrt{\frac{k^2}{q^2}}-1\right)
\Theta(1-q^2/k^2),
\end{equation}
we obtain $\ell_1^{(\mathrm F)}=2/3$. By another rescaling $g
\mapsto g \pi^2/4\ell_1^{(\mathrm F)}$ and $\tilde g \mapsto \tilde g
\pi^2/4\ell_1^{(\mathrm F)}$,
this multiplicative regulator dependence drops out,
\begin{align}
\label{eq:beta_tilde-g_dimless}
\partial_t \tilde g & = \tilde g - \frac{2\Nf-1}{2\Nf} \tilde g^2 +
\frac{3}{2\Nf} \tilde g g + \frac{1}{\Nf} g^2, \\
\label{eq:beta_g_dimless}
\partial_t g & = g + \frac{1}{2\Nf} \tilde g g + \frac{2\Nf + 1}{6\Nf}
g^2.
\end{align}
For $\Nf=1$, our result coincides with the result found via a perturbative RG
approach \cite{Herbut:2009qb} in the context of interacting electrons on the
honeycomb lattice.

\section{Fixed points and critical exponents} \label{sec:fixed-points}
From the coupling flows, it is straightforward to analyze the fixed-point
structure in order to study possible asymptotically safe UV trajectories of the
RG flow. A fixed point $g^\ast$ is defined by
\begin{equation}
\forall i: \quad \beta_i(g_1^\ast,g_2^\ast,\dots) = 0,
\end{equation}
with $\beta_i \equiv \partial_t g_i$. Whereas the fixed-point values are
regulator-scheme dependent, see above, the linearized flow in the vicinity of
the fixed point is universal as quantified by the critical exponents. More
specifically, the Jacobian $B_i{}^j$ of the linearized flow near a fixed point,
\begin{equation}
\partial_t g_i = B_i{}^j(g_j - g_j^\ast) + \dots, \qquad B_i{}^j =
\left.\frac{\partial \beta_i}{\partial g_j}\right\vert_{g=g^\ast},
\end{equation}
defines the {\em stability matrix}. The associated eigenvectors $v$ govern the
evolution of small deviations from the fixed point according to $\partial_t v
= B v = - \Theta v$. The corresponding eigenvalues (including a minus sign)
$\Theta$ are universal and can be associated with thermodynamic critical
exponents if the fixed point corresponds to a critical point of a 2nd order
phase transition. For brevity, all $\Theta$'s are referred to as critical
exponents.  The solution to the linearized flow $v \propto k^{-\Theta}$
implies that positive $\Theta>0$ correspond to RG relevant, i.e., infrared
repulsive, directions and negative $\Theta<0$ correspond to RG irrelevant,
i.e., infrared attractive, directions. 

For the present set of flow equations, a general property of the fixed points
and their critical exponents can be proven \cite{Gies:2003dp}: the beta
functions all are of the form
\begin{equation}
\beta_i = g_i + g_k A^{kl}_i g_l \quad \Rightarrow \quad B_i{}^j = \delta_i{}^j
+ 2 g^\ast_k A_i^{kj},
\end{equation}
with matrices $A_i^{kl}$ which are symmetric in the upper indices. We now see
that for every interacting fixed point $g^\ast\neq 0$ the fixed-point vector
itself is an eigenvector of $B$, $v=g^\ast$, with the critical exponent
$\Theta = 1$,
\begin{equation}\label{eq:int_fp_rel-direc}
B_i{}^j g_j^\ast = g_i^\ast + 2 g_k^\ast A_i^{kj} g_j^\ast = - g_i^\ast.
\end{equation}
Here, we have made use of the fixed point equation $\beta_j(g^\ast) = 0$. We
conclude that every fixed point besides the Gau\ss ian fixed point $g^\ast=0$
has at least one relevant and thus infrared repulsive direction. Each
non-Gau\ss ian fixed point is therefore a candidate for a possible UV
completion, potentially defining an own universality class
\cite{Gies:2003dp}. For the Gau\ss ian fixed point $g^\ast=0$, the stability
matrix is just the identity $B_i{}^j = \delta_i{}^j$, such that the Gau\ss ian
fixed point is infrared attractive in every direction with $\Theta=-1$, giving
rise to only trivial theories at long ranges.

\begin{figure}[tb]
\includegraphics[width=.48\textwidth]{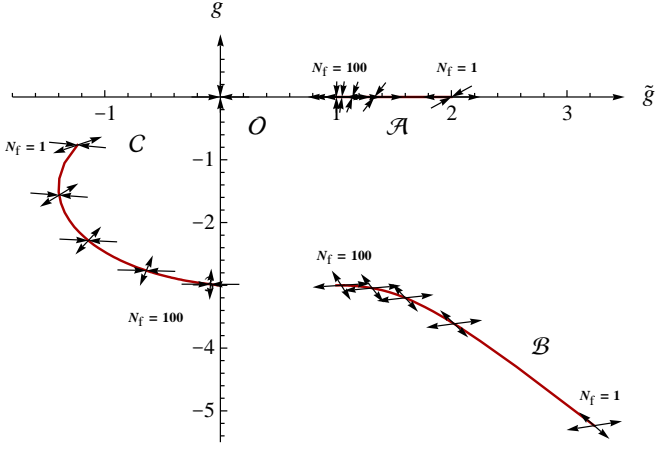}
\caption{\label{fig:lines-fixpt} Positions of RG fixed points and
  \mbox{(ir-)relevant} directions for flavor numbers $\Nf=1,2,4,10,100$.
  Arrows denote the RG flow towards the IR. In agreement with
  Eq.~\eqref{eq:int_fp_rel-direc}, one relevant direction of every interacting
  fixed point points towards the Gau\ss ian fixed point $\mathcal O$.}
\end{figure}

For fixed $g_{j\neq i}$, the beta function $\beta_i$ corresponds graphically
to a parabola, such that we expect for our truncation exactly $2^2=4$
(possibly complex or degenerate) solutions of the fixed-point equations. For
any $\Nf\in\mathbb N$, we find them to be real and non-degenerate;
the explicit solutions $(\tilde g^\ast,g^\ast)$ for the Gau\ss ian fixed point
$\mathcal O$ and the three non-Gau\ss ian fixed points $\mathcal{A,B,C}$ are
\begin{align}
\mathcal O&: \ \left(0, 0\right)\\
\mathcal A&: \ \left( \frac{2\Nf}{2\Nf-1}, 0\right)
\label{eq:fixed-point_A} \displaybreak[0] \\
\mathcal B&: \ \left(\frac{\Nf\left(14 - 7\Nf + 2\Nf^2 + (1+2\Nf)\sqrt{16 +
28 \Nf + \Nf^2}\right)}{-5 + 8\Nf + 2\Nf^2 + 4\Nf^3},\right.\notag \\
&\quad \left.-\frac{12\Nf^2}{4+4\Nf+4\Nf^2 - \sqrt{16+28\Nf+\Nf^2}}\right)
\label{eq:fixed-point_B} \displaybreak[0] \\
\mathcal C&: \ \left(\frac{\Nf\left(14 - 7\Nf + 2\Nf^2 - (1+2\Nf)\sqrt{16 +
28 \Nf + \Nf^2}\right)}{-5 + 8\Nf + 2\Nf^2 + 4\Nf^3},\right.\notag \\
&\quad \left.-\frac{12\Nf^2}{4+4\Nf+4\Nf^2 + \sqrt{16+28\Nf+\Nf^2}}\right).
\label{eq:fixed-point_C}
\end{align}
It is straightforward to derive the eigenvectors and eigenvalues of the
stability matrix (and thus the critical exponents and the RG
\mbox{(ir-)relevant directions}) analytically; but as the general formulas may
not provide much physical insight, we present the results graphically: in
Fig.~\ref{fig:lines-fixpt}, we plot the positions of the fixed points in
theory space (spanned by the two couplings $\tilde g$ and $g$) together with the
corresponding eigenvectors of the stability matrix $B_i{}^j$ for various flavor
numbers $\Nf$.
%
\begin{figure}[tb]
\includegraphics[width=.48\textwidth]{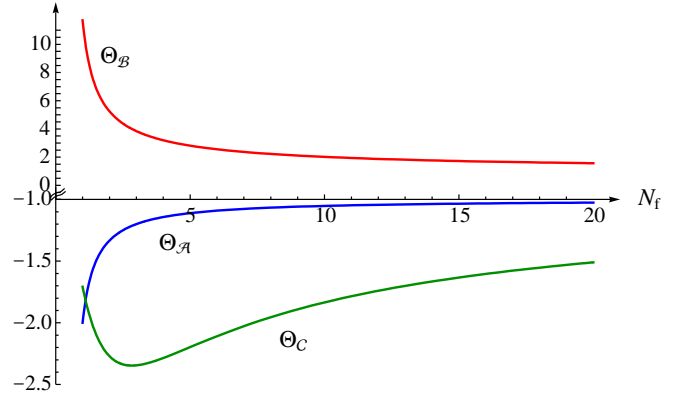}
\caption{\label{fig:crExpPlot} Critical exponents $\Theta$ as a function of
  flavor number $\Nf$ for the interacting fixed points $\mathcal A$, $\mathcal
  B$, $\mathcal C$. The value of the corresponding other critical exponent is
  $\Theta=1$ for all $\Nf$.  Note that the scale of the vertical axis changes
  at $\Theta=-1$.}
\end{figure}
%
In Fig.~\ref{fig:crExpPlot}, the (non-trivial) critical exponents $\Theta$ for
the interacting fixed points are given as a function of $\Nf$ . For any $\Nf$,
fixed point $\mathcal B$ has two RG relevant directions, whereas the
interacting fixed points $\mathcal A$ and $\mathcal C$ have one relevant and
one irrelevant direction. As the number of relevant directions corresponds to
the number of physical parameters to be fixed, theories emanating from
$\mathcal A$ and $\mathcal C$ are fully determined, once the initial condition
for this relevant direction is fixed. In the sense of dimensional
transmutation, fixing this one parameter can be viewed as fixing a total scale
for the system. Therefore, theories belonging to these universality classes
defined by $\mathcal A$ and $\mathcal C$ are fully predictive, once a global
scale is fixed. Theories emanating from $\mathcal B$ are fixed by a mass scale
and one further parameter, e.g., a dimensionless coupling ratio, whereas
$\mathcal O$ does not support an interacting system at long ranges. 

At this point, let us already stress that no interacting fixed point is on the
pure Thirring axis ($\tilde g=0$) for any finite $\Nf$. In fact, as 
discussed below, we associate the Thirring universality class with fixed point
$\mathcal C$ which approaches the pure Thirring coupling only in the
asymptotic limit $\Nf\to \infty$. For any finite $\Nf$, the renormalized UV
trajectory of the Thirring model will have to pass through the full
two-dimensional coupling plane, even though the long-range physics does depend
only on one physical parameter (e.g., the value of the Thirring coupling at a
certain scale).

\section{Phase transitions and long-range physics} \label{sec:discussion}
A technical means for the discussion of long-range phases are the
separatrices, i.e., those RG trajectories that interpolate between two fixed
points. They subdivide the theory space into separate flow regions, providing a
classification which can potentially be related to spontaneous symmetry
breaking in the long-range limit. At this point, we stress that the fermionic
truncation is not sufficient for a complete discussion of long-range physics
which is expected to be dominated by composite bosonic degrees of freedom such
as condensates and excitations on top of condensates. 

\begin{figure}[tb]
\includegraphics[width=.48\textwidth]{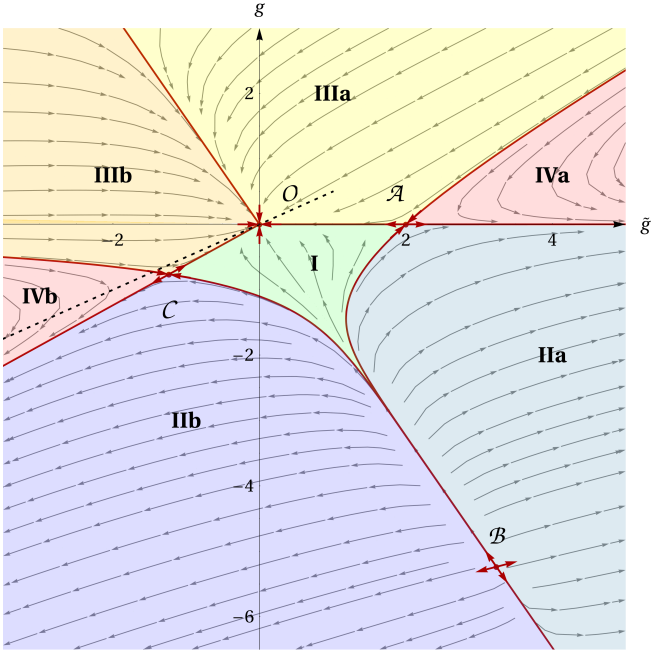}
\caption{\label{fig:flow1-flavor} Classification of Thirring-like 4-fermi
  theories determined by the fixed-point positions and the corresponding RG
  trajectories (arrows denote the flow towards the IR). Red (solid) lines
  depict separatrices that interpolate between fixed points and separate
  different regions. The vertical axis $\tilde g=0$ corresponds to
  models with a pure Thirring coupling. The dashed line ($g=\tilde g/2$) marks
  a generalized NJL model, see text.}
\end{figure}

Let us start with a closer look at the one-flavor case $\Nf=1$. The
separatrices subdivide the theory space into distinct sections defined by
their IR and UV behavior; in Fig.~\ref{fig:flow1-flavor}, we plot the RG flow
using Eqs.~\eqref{eq:beta_tilde-g_dimless}--\eqref{eq:beta_g_dimless}.  A
classification of RG trajectories is listed in Tab.~\ref{tab:theory_classes}.
The IR behavior of the theories in the regions I, IIIa and IIIb is governed by
the Gau\ss ian fixed point $\mathcal O$. In these regions, both couplings
$\tilde g$ and $g$ are irrelevant, leading to non-interacting theories in the
IR. The regions IIa and IVa are characterized by an irrelevant Thirring
coupling $g$, $\lim_{k\rightarrow 0} g = 0$. We thus expect that the bosonic
channel $S\sim \bar\psi^a \gamma_{45} \psi^a$ becomes critical at a
sufficiently large $\tilde g$, dynamically generating a parity breaking mass
$\tilde m \propto \langle\bpsi \gamma_{45} \psi\rangle$. If so, the fixed
point $\mathcal A$ governs the spontaneous breaking of parity, potentially
being associated with a 2nd order phase transition defining a new universality
class.

\begin{table}[tb]
\caption{\label{tab:theory_classes} Classification of all RG trajectories,
c.f.\ Fig.~\ref{fig:flow1-flavor}.}
\begin{ruledtabular}\begin{tabular}{cccc}
Type & UV behavior & IR behavior & universality class\\ 
 & $\lim_{k\rightarrow 0}(\tilde g, g)$ & $\lim_{k\rightarrow
\Lambda}(\tilde g, g)$ \\ \hline
I & $\mathcal B$ & $\mathcal O$ &  non interacting\\
I-IIa separ. & $\mathcal B$ & $\mathcal A$ \\
I-IIb separ. & $\mathcal B$ & $\mathcal C$ \\
I-IIIa separ. & $\mathcal A$ & $\mathcal O$ & non interacting \\
I-IIIb separ. & $\mathcal C$ & $\mathcal O$ & non interacting \\
IIa & $\mathcal B$ & $(\infty,0)$ & parity breaking \\
IIa-IIb separ. & $\mathcal B$ & $\mathcal B \cdot \infty$ \\
IIa-IVa separ. & $\mathcal A$ & $(\infty, 0)$ & parity breaking \\
IIb & $\mathcal B$ & $\mathcal C \cdot \infty $ & chirality breaking \\
IIb-IVb separ. & $\mathcal C$ & $\mathcal C \cdot \infty $ & chirality breaking
\\
IIIa & $(-\mathcal C) \cdot \infty $ & $\mathcal O$ & non interacting \\
IIIa-IIIb separ. & $(-\mathcal B) \cdot \infty$ & $\mathcal O$ & non interacting
\\
IIIa-IVa separ. & $(-\mathcal C) \cdot \infty $ & $\mathcal A$ \\
IIIb & $(-\infty,0)$ & $\mathcal O$ & non interacting \\
IIIb-IVb separ. & $(-\infty,0)$ & $\mathcal C$\\
IVa & $(-\mathcal C) \cdot \infty $ & $(\infty, 0)$ & parity breaking \\
IVb & $(-\infty,0)$ & $\mathcal C \cdot \infty$ & chirality breaking
\end{tabular}\end{ruledtabular}
\end{table}

By contrast, both $\tilde g$ and $g$ diverge in the regions IIb and IVb in the
infrared limit. To interpret this behavior, let us (for arbitrary $\Nf\geq 1$)
rewrite the interaction terms by means of the Fierz theorem as
\begin{multline}\label{eq:thirring-NJL-interaction}
\frac{\tilde g}{2\Nf}S^2 + \frac{g}{2\Nf}S_\mu^2 = 
\frac{1}{4\Nf}\left\{(2 g - \tilde g)(\bpsi^a\gamma_\mu\psi^a)^2 \right.\\
 - \tilde g\left[(\bpsi^a\psi^a)^2 - (\bpsi^a\gamma_4\psi^a)^2 -
(\bpsi^a\gamma_5\psi^a)^2\right] \\
\left. + \tilde g \left[(\bpsi^a\gamma_{45}\psi^a)^2 -
(\bpsi^a\gamma_{45}\psi^b)^2\right]\right\},
\end{multline}
cf. Eqs.~\eqref{eq:invariant_4-fermi}--\eqref{eq:invariant_4-fermi_fierz-B}
and App.~\ref{app:fierz}.  In the one-flavor case $\Nf = 1$, the last term
(proportional to $+\tilde g$) vanishes. The interaction then simply is a
linear combination of the Thirring interaction with coupling $2g- \tilde g$
and a generalized Nambu-Jona-Lasinio (NJL)-type \cite{Nambu:1961tp}
interaction with coupling $-\tilde g$.  Along the straight line through
$\mathcal O$ and $\mathcal C$ which may govern the IR behavior of the
trajectories IIb and IVb the NJL-type interaction in fact dominates, $|\tilde
g|/|2g-\tilde g| \approx 4.24 \gg 1$. To illustrate this, we have also plotted
in Fig.~\ref{fig:flow1-flavor} the line of vanishing Thirring interaction $2g
- \tilde g = 0$ in the Fierz-transformed form as a dashed line. This resulting
NJL-type line is fairly close to the separatrix through $\mathcal O$ and
$\mathcal C$ (red line). We thus expect that at a large coupling on this
separatrix the NJL-type channel eventually becomes critical, inducing a
symmetry-broken state with
\begin{equation*}
m \propto \langle \bpsi \psi \rangle \neq 0.
\end{equation*}
This is equivalent to saying that $\langle \bpsi \gamma_4 \psi \rangle \neq 0$
or $\langle \bpsi \gamma_5 \psi \rangle \neq 0$ is expected to be favored in
this state.
For $\Nf = 1$, we therefore identify the fixed point $\mathcal C$ with the
critical point governing the phase transition into the chiral symmetry broken
phase for all theories of the regions IIb and IVb, in agreement
with~\cite{Herbut:2009qb}. This is precisely the behavior which is expected in
the Thirring model, hence we associate all trajectories emanating from $\mathcal
C$ with UV complete fully renormalized versions of the 3$d$ Thirring model. 

\begin{figure}[tbp]
\includegraphics[width=.48\textwidth]{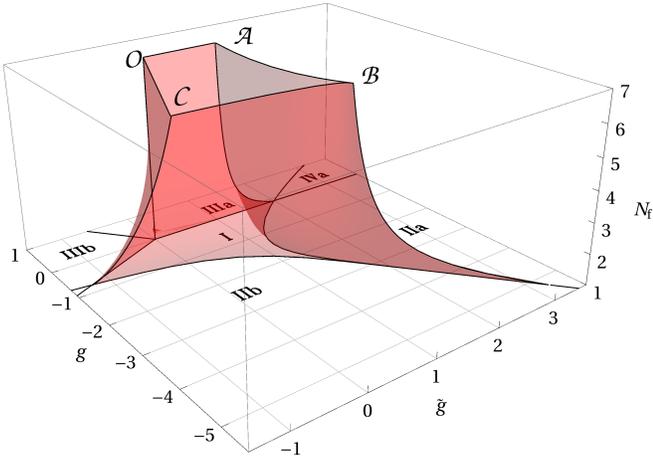}
\caption{\label{fig:crit-surface} Fixed-point positions and separatrices in
  the ($g,\tilde g$) coupling plane as a function of flavor number for $1\leq
  \Nf\leq 7$. The horizontal slice at $\Nf=1$  is equivalent to
  Fig.~\ref{fig:flow1-flavor}.}
\end{figure}

Provided the last term in Eq.~\eqref{eq:thirring-NJL-interaction} proportional
to $+\tilde g$ does not dominate even for higher number of flavors, we may
extend the preceding discussion to larger $\Nf\geq 1$.  In
Fig.~\ref{fig:crit-surface}, we show how the positions of the fixed points and
the separatrices behave for increasing flavor number. For $\Nf \rightarrow
\infty$ the region I turns into a rectangle with the vertices $(\tilde g, g) =
(0,0)$ and $(1,-3)$. The Thirring fixed point $\mathcal C$ hence moves to
$(0,-3)$ and becomes a pure Thirring coupling, c.f.\ also
Fig.~\ref{fig:lines-fixpt}. Thus, the IR attractive line
$\mathcal{OC}$ approaches the Thirring axis $\tilde g = 0$ for
larger flavor number.

Along this line of coupling values, we expect the flow to be dominated by the
vector channel $\bar\psi^a \gamma_\mu \psi^a$ which agrees precisely with the
dominant bosonic degree of freedom in a large-$\Nf$ analysis
\cite{Parisi:1975im, Hands:1994kb, Gomes:1990ed, Hong:1993qk}. As
there is no chiral symmetry breaking at large $\Nf$, it is natural to expect a
quantum phase transition to occur for increasing $\Nf$ while the line
$\mathcal{OC}$ undergoes a transition from the NJL-regime $g\sim \tilde g/2$
to the large-$\Nf$ regime where $\tilde g=0$. Unfortunately, a more
quantitative picture of this quantum phase transition is difficult to obtain
in the purely fermionic language. A quantitative RG analysis requires the
inclusion of dynamical chiral (i.e., NJL-type) and vector bosonic degrees of
freedom in order to study the interplay of these competing orders as a
function of $\Nf$. This is left for future work. 

For the remainder of this section, we shall be satisfied with a simple
estimate of the transition region. As a rough criterion, let us determine the
flavor number where the separatrix $\mathcal{OC}$ is half way in-between the
generalized NJL model on the one hand and the pure Thirring coupling on the
other hand. This value of $\Nf$ follows from 
\begin{equation}
\left.\frac{\left|\tilde g\right|}{\left|2g-\tilde g\right|}\right|_\mathcal C
= \frac{3}{\Nf - 7 + \sqrt{16 + 28\Nf + \Nf^2}} \approx 1,
\end{equation}
and is given by $\Nf \approx 7/4$. We stress that this number should not be
viewed as a direct estimate of $\Nfc$, as the onset of true critical behavior
can easily provide further correction factors of order $\mathcal{O}(1)$.  

\begin{figure}[tbp]
\includegraphics[width=.48\textwidth]{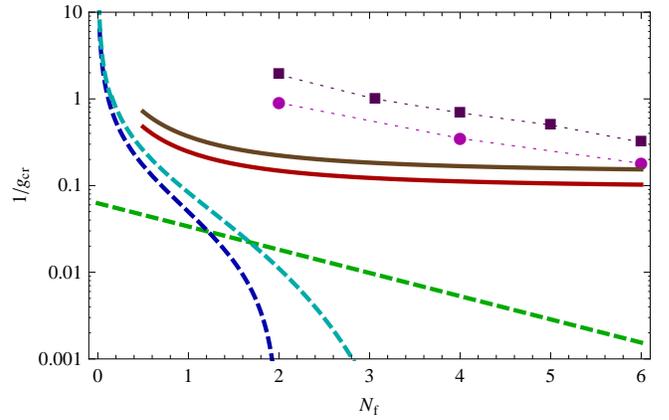}
\caption{\label{fig:crit-coupling} Comparison of non-universal critical
  Thirring couplings from different methods. Solid lines display the critical
  couplings from this work taken as the $g$ coordinate of the intersection
  point of the separatrix~$\mathcal BC$ with the Thirring axis
  $\tilde g = 0$ for the sharp cutoff (upper solid/brown line) or the linear
  regulator (lower solid/red line). Dotted: Monte Carlo results (magenta
  with circles: \cite{Kim:1996xza}, purple with squares:
  \cite{Christofi:2007ye}). Dashed: DSE approaches (from right to left: green
  \cite{Hong:1993qk}, cyan \cite{Itoh:1994cr} and blue \cite{Kondo:1995jn}).}
\end{figure}

It is instructive to compare our results with those obtained by other
methods such as Monte Carlo simulations or truncated DSEs. A variety of
studies have computed the modulus of the critical Thirring coupling which is a
necessarily required for chiral symmetry breaking (but not sufficient beyond
the critical flavor number). Two cautionary remarks are in order: first of
all, these critical couplings similarly as the fixed-point values are not
universal, such that the choice of the regularization scheme can have a
strong quantitative influence on the estimated values. Second, most other
studies have defined the microscopic Thirring model exactly on the Thirring
axis $\tilde g=0$; in principle, the full coupling plane has to be considered
such that the critical coupling on the axis may be different (larger in
modulus) from a corresponding estimate directly at the fixed point. 

To circumvent the second caveat, we also consider the coupling on the Thirring
axis $\tilde g = 0$ with an initial value of the coupling $g$ above the
separatrix which interpolates between the fixed points $\mathcal B$ and
$\mathcal C$. Then the theory is in the region I or IIIa, depending on the
sign of $g$, such that the RG flow drives the couplings to the free theory at
Gau\ss ian fixed point $\mathcal O$. Therefore, the absolute value of the $g$
coordinate of the intersection point of this separatrix with the Thirring axis
provides a lower bound for the absolute value of the critical coupling
$g_\text{cr}$ at which the $\chi$SB phase transition occurs. We compare this
lower bound for different regulator functions $R_k$ with the values of the
critical coupling obtained by Monte Carlo simulations \cite{Kim:1996xza,
  Christofi:2007ye} and different sequences of truncated DSE approaches
\cite{Hong:1993qk, Itoh:1994cr, Kondo:1995jn}. In Fig.~\ref{fig:crit-coupling},
we plot this inverse coupling $1/g_\text{cr}$ for varying number of flavors
$0<\Nf\leq 6$. For the sharp cut-off as well as for the linear cut-off our
results lie well above the values obtained by DSE approaches but below the
values from lattice-regularized Monte Carlo simulations.  Note that similar to
the lattice and one of the DSE results, we do not observe a sharp decay of the
inverse critical coupling. This behavior would indicate a sharp growth of
$g_\text{cr}$ above a certain number of flavors, corresponding to a critical
flavor number in the infinite coupling limit. Instead, we observe a rather
smooth dependence on $\Nf$ as on the lattice for $\Nf\lesssim 6$, which is
compatible with our expectation that the quantum phase transition towards the
chirally symmetric phase occurs because of competing large-$\Nf$ degrees of
freedom  and not because of a change in the UV critical structure.

\section{Conclusions} \label{sec:conclusions}
In this work, we have investigated 3$d$ relativistic fermionic models in a 
theory space, defined by chiral and a set of discrete symmetries and
point-like interactions. Even though the construction of these models has been
inspired by the uniquely fixed 2$d$ Thirring model, the corresponding 3$d$
symmetries involving a reducible 4-component Dirac spinor representation
enlarge the minimal coupling space and give room for a larger fixed-point
structure and thus for different microscopic realizations of such fermion
models.

We have classified all point-like interactions satisfying the symmetry
constraints and determined their RG flow in a systematic next-to-leading order
derivative expansion. The fact that leading-order and next-to-leading order
results are identical as the anomalous dimension remains zero can be
interpreted as a signature for the convergence of the expansion, as long as no
further composite channels develop a strong RG flow. 

The resulting flow equations for the two independent fer\-mionic couplings
generate a fixed-point structure of one trivial Gau\ss ian and three
interacting fixed points $\mathcal{A,B,C}$ which can be classified by their
critical exponents. We associate all RG trajectories emanating from the fixed
point $\mathcal C$ with fully renormalized UV complete versions of the 3$d$
Thirring model; as this fixed point has one RG relevant direction, fixing one
physical scale suffices to obtain a fully IR predictive quantum field theory
in the Thirring universality class. 

Fixed point $\mathcal A$ which, incidentally, lives in an RG invariant
sub-manifold of theory space (defined by $g=0$) is also characterized by only
one RG relevant direction. From the nature of the scalar channel associated
with this coupling direction, we conjecture that this fixed point can be
related to a critical point of a 2nd order phase transition to a phase with
broken parity. The third interacting fixed point $\mathcal B$ has two relevant
directions; depending on the initial conditions, theories emanating from this
fixed point can flow to the Thirring phase as well as to the parity-broken or
symmetric phases.  

Unfortunately, our purely fermionic RG analysis does so far not permit us to
reliably run towards or into the symmetry-broken phases. Such a quantitative
description is required in order to analyze the true IR behavior of the
Thirring phase. From the structure of the fermionic flow, in particular, from
the behavior of the separatrix $\mathcal{OC}$, we conjecture that
the long-range dynamics is characterized by a competition between NJL-type
chiral condensation channels on the one hand and large-$\Nf$-type vector
bosons on the other hand. As the vector-boson fluctuations generically inhibit
chiral symmetry breaking, we expect the occurrence of a quantum phase
transition of the Thirring model at a critical flavor number $\Nfc$,
separating a broken phase for small $\Nf$ from a disordered phase for large
$\Nf$. Our very rough estimate of the transition region in any case is
compatible with the findings from lattice simulations indicating that $\Nfc
\simeq 6.6$~\cite{Christofi:2007ye}.

Our results on the position of the Thirring fixed point $\mathcal C$ being
actually away from the pure Thirring axis $\tilde g=0$ for any finite $\Nf$
provokes an important comment: both lattice simulations as well as DSE studies
build on a microscopic definition of the 3$d$ Thirring model which is fixed
only with the Thirring coupling, i.e., by a pure bare Thirring-like action. Our
fixed-point results indeed provide a fundamental justification for this, as
the Thirring fixed point indeed is characterized by only one relevant
direction. As long as the microscopic actions chosen in other formulations are
in a sufficiently attractive domain of the Thirring fixed point, universality
guarantees that the long-range physics is indeed purely governed by the
Thirring fixed point. We expect this to hold also for the determination of the
critical flavor number (there is no universality for quantities such as the
scheme-dependent critical coupling, see Fig.~\ref{fig:crit-coupling}).
Nevertheless, one caveat should be emphasized: this conclusion about
universality only holds, as long as the microscopic bare actions indeed are in
the attractive domain of the Thirring fixed point. For instance, if by
accident a lattice formulation turned out to be influenced by fixed point
$\mathcal B$, the simulation would simply describe a model different from the
Thirring model potentially exhibiting a different quantum phase transition as
a function of $\Nfc$. As the fixed-point positions are not universal, our
results are unfortunately not directly transferable to the lattice theory
space. However, provided that the fixed-point structure is qualitatively
similar, our results can be taken as a support for the implicit assumption
that the lattice simulations have indeed been performed in the real Thirring
universality class.

Given the importance of the quantitative value of the critical flavor number
$\Nfc$ of the Thirring model in the light of condensed-matter applications, a
natural next step of our studies will be the inclusion of composite degrees of
freedom in order to study the competition among the various bosonic
channels. Within the functional RG, this can conveniently be investigated by
means of partial or dynamical bosonization \cite{Gies:2001nw}. As the problem
of competing order is a paradigmatic one in (quantum) critical phenomena and
statistical physics \cite{Salmhofer:2004}, we consider the relativistic 3$d$
Thirring model as an ideal test case.

\acknowledgments{Helpful discussions with J.~Braun, S.~Hands, S.~Rechenberger,
  D.~Scherer, and M.~Scherer are gratefully acknowledged.  This work has been
  supported by the DFG under GRK 1523, FOR 723 and Gi 328/5-1.}

\appendix
\section{Fierz transformations} \label{app:fierz}
Let $(\gamma^A , \gamma^B) \coloneqq \Tr(\gamma^A\gamma^B)$
define an inner product on the space of $4\times 4$ Dirac matrices. The $16$
matrices $\{\gamma^A\}$ in Eq.~\eqref{eq:basis_dirac} are orthogonal with
respect to this product,
\begin{equation}
\Tr(\gamma^A\gamma^B) = 4 \delta^{AB},
\end{equation}
and thus define a complete basis of the $4\times 4$ Dirac matrices (which
generically have $16$ independent matrix elements),
\begin{equation} \label{eq:Dirac_completeness}
\sum_{A=1}^{16} \frac{1}{4}\gamma^A_{m\ell}\gamma^A_{ik} =
\delta_{mk}\delta_{i\ell}.
\end{equation}
The Fierz transformations are straightforwardly obtained by multiplying the
completeness relation \eqref{eq:Dirac_completeness} by each of the 4-fermi terms
$\bpsi^a_m(\gamma_A \psi^b)_k \bpsi^c_i (\gamma_A\psi^d)_\ell$ and, where
appropriate, decomposing multiple products of Dirac matrices into basis
elements $\gamma_A$. We find
\begin{equation}
\left(\bpsi^a \gamma_A \psi^b\right)\left(\bpsi^c \gamma_A \psi^d\right) =
\sum_{B=1}^{16} C_{AB} \left(\bpsi^a \gamma_B \psi^d\right)\left(\bpsi^c
\gamma_B \psi^b\right)
\end{equation}
with
\begin{equation}
(C_{AB}) = \frac{1}{4}
\begin{pmatrix}
-1&-1&-1&-1&-1&-1&-1&-1 \\
-3&1&3&1&-1&-1&-3&3 \\
-1&1&-1&-1&1&-1&1&1 \\
-3&1&-3&1&1&1&-3&-3 \\
-3&-1&3&1&1&-1&3&-3 \\
-3&-1&-3&1&-1&1&3&3 \\
-1&-1&1&-1&1&1&-1&1 \\
-1&1&1&-1&-1&1&1&-1
\end{pmatrix}
\end{equation}
and $(\gamma_A) = (\mathbbm 1,\gamma_\mu,\gamma_4,
\frac{\sigma_{\mu\nu}}{\sqrt{2}},\iu \gamma_\mu\gamma_4,\iu \gamma_\mu\gamma_5,
\gamma_{45},\gamma_5)^\mathrm T.$
With these preliminaries one can simply read off the Fierz relations between the
invariant 4-fermi interactions in Eqs.~\eqref{eq:invariant_4-fermi},
\eqref{eq:invariant_4-fermi_fierz-A} and \eqref{eq:invariant_4-fermi_fierz-B}.

\section{Derivation of flow equations}\label{app:flow_eqs}
Consider the right-hand side of the flow
equation \eqref{eq:wetterich-eq}. By introducing the derivative $\tilde
\partial_t$, acting per definition only on the $t$ dependence of the regulator
$R_k$, we may expand
\begin{align}
\frac{\partial_t R_k}{\Gamma^{(2)}_k+R_k}& = 
\tilde\partial_t \ln \left(\Gamma_{k,0}^{(2)} + R_k +
\Delta\Gamma_k^{(2)}\right) \\
& = \tilde\partial_t \ln \left(\Gamma_{k,0}^{(2)} + R_k\right)
+ \tilde\partial_t \left(\frac{\Delta\Gamma_k^{(2)}}{\Gamma_{k,0}^{(2)} +
R_k}\right) \notag\\ \label{eq:log-expansion}
&\quad - \frac{1}{2}\tilde\partial_t
\left(\frac{\Delta\Gamma_k^{(2)}}{\Gamma_{k,0}^{(2)}+R_k}\right)^2 
+ \dots
\end{align}
with $\Delta\Gamma^{(2)}_k$ containing the field-dependent
parts of $\Gamma^{(2)}_k$, and $\Delta\Gamma^{(2)}_{k,0}$ containing the
field-independent (propagator) part, such that $\Gamma^{(2)}_k =
\Gamma^{(2)}_{k,0} + \Delta\Gamma^{(2)}_k$.
The Hessian of the average effective action in terms of Fourier transformed
fields $\psi(q) \equiv \psi_q$ and $\bpsi(q)\equiv \bpsi_q$ is given by
\begin{equation}
\Gamma_k^{(2)}(p,q) =
\begin{pmatrix}
  \frac{\overrightarrow\delta}{\delta \psi^a_{-p}{}^\mathrm T}\Gamma_k
  \frac{\overleftarrow\delta}{\delta \psi^b_q}
& \frac{\overrightarrow\delta}{\delta \psi^a_{-p}{}^\mathrm T}\Gamma_k
  \frac{\overleftarrow\delta}{\delta \bpsi^b_{-q}{}^\mathrm T} \\
\frac{\overrightarrow\delta}{\delta \bpsi^a_p}\Gamma_k
  \frac{\overleftarrow\delta}{\delta \psi^b_q}
& \frac{\overrightarrow\delta}{\delta \bpsi^a_p}\Gamma_k
  \frac{\overleftarrow\delta}{\delta \bpsi^b_{-q}{}^\mathrm T}
\end{pmatrix}.
\end{equation}
The corresponding fluctuation matrix results in\\
\begin{widetext}\begin{align}
\Delta\Gamma^{(2)}_k(p,q) & = \frac{\tilde{\bar{g}}_k}{\Nf} 
  \begin{pmatrix}
    - \int_{p_1}\left(\bpsi^a_{p_1} \gamma_{45}\right)^\mathrm T
    \left(\bpsi^b_{q-p-p_1}\gamma_{45}\right)
  & \int_{p_1} \left\{\left(\bpsi^a_{p_1}\gamma_{45}\right)^\mathrm T
    \left(\gamma_{45} \psi^b_{p-q+p_1}\right)^\mathrm T -
    \left(\bpsi_{p_1}\gamma_{45}\psi_{p-q+p_1}\right) 
    \gamma_{45}^\mathrm T \delta^{ab} \right\} 
  \\
    \int_{p_1}\left\{\left(\bpsi_{p_1}\gamma_{45}\psi_{p-q+p_1}\right)
    \gamma_{45} \delta^{ab} +
    \left(\gamma_{45}\psi^a_{p_1}\right) \left(\bpsi^b_{q-p+p_1}
    \gamma_{45}\right)\right\}
  & - \int_{p_1} \left(\gamma_{45}\psi^a_{p_1}\right)
    \left(\gamma_{45}\psi^b_{p-q-p_1}\right)^\mathrm T
  \end{pmatrix} \notag \\
& \quad + \text{term with } (\tilde{\bar{g}}_k,\gamma_{45})
  \leftrightarrow (\bar{g}_k, \gamma_\mu)
\end{align}\end{widetext}
with $\int_{p_1} \equiv \int \frac{\mathrm{d}^3 p_1}{(2\pi)^3}$, and for the
propagator
\begin{equation}
\left(\Gamma_{k,0}^{(2)} + R_k\right)^{-1} = - \frac{ (2\pi)^3
\delta^{(3)}_{p-q}\delta^{ab}}{Z_k q^2 (1 + r(q^2/k^2))}
\begin{pmatrix}
  0 & \slashed{q} \\
  \slashed{q}^\mathrm T & 0
\end{pmatrix},
\end{equation}
with $R_k(q) = Z_k \slashed{q} r(q^2/k^2)$. 

\paragraph*{Anomalous dimension.} The flow of the wave function renormalization
$Z_k$ is obtained by a suitable projection of the Wetterich equation
\eqref{eq:wetterich-eq},
\begin{equation}
\partial_t Z_k = 
\frac{1}{24} \Tr \left\{ \left.
  \gamma_\mu \frac{\partial}{\partial p'_\mu}\int_{q'}
  \frac{\overrightarrow\delta}{\delta\bpsi^1_{p'}}
  \left[\Tr\left(\frac{\partial_t R_k}{\Gamma_k^{(2)} + R_k}\right)\right]
  \frac{\overleftarrow\delta}{\delta\psi^1_{q'}}
\right \vert_{\psi=\bpsi=0} \right \} .
\end{equation}
Using the expansion \eqref{eq:log-expansion}, we observe that only the term
linear in $\Delta \Gamma_k^{(2)}$ survives the projection. Since this term is
traceless, the flow of the wave function renormalization in this order of the
expansion vanishes,
\begin{equation}
\eta_k = -\partial_t \ln Z_k \equiv 0.
\end{equation}
This line of argument is reminiscent to a similar observation of the vanishing
of the anomalous dimension in scalar $O(N)$ models in the symmetric regime at
next-to-leading order derivative expansion. 

\paragraph*{4-fermi couplings.} Obviously only the term in
Eq.~\eqref{eq:log-expansion} being quadratic in $\Delta\Gamma_k^{(2)}$
leads to 4-fermi terms on the right-hand side of the Wetterich
equation \eqref{eq:wetterich-eq},  contributing to the flow of the
\mbox{4-fermi} couplings.
Evaluating this term for constant fields and taking the trace over flavor,
spinor, and momentum degrees of freedom, we infer\\
\begin{widetext}\begin{multline} \label{eq:trace_gamma2}
\frac{1}{2} \Tr \left[\frac{1}{2}\tilde\partial_t
\left(\frac{\Delta\Gamma_k^{(2)}}{\Gamma_{k,0}^{(2)}+R_k} \right)^2\right] = \\
\frac{2\Omega}{\Nf\pi^2}
  \Intdq \tilde \partial_t \frac{1}{Z_k^2\left[1+r(q^2/k^2)\right]^2}
    \left[\left(\frac{2\Nf-1}{2\Nf} \tilde{\bar{g}}_k^2 
    - \frac{3}{2\Nf}\bar{g}_k\tilde{\bar{g}}_k 
    - \frac{1}{\Nf} \bar{g}_k^2\right) (\bpsi\gamma_{45}\psi)^2
  + \left(-\frac{1}{2\Nf} \bar{g}_k \tilde{\bar{g}}_k
    - \frac{2\Nf + 1}{6\Nf} \bar{g}_k^2\right) (\bpsi\gamma_\mu\psi)^2\right],
\end{multline}\end{widetext}
with $\Omega$ being the spacetime volume. The desired beta functions for
$\tilde{\bar{g}}_k$ and $\bar{g}_k$, respectively, are simply given by
$2\Nf/\Omega$ times the coefficient of the corresponding 4-fermi term in
Eq.~\eqref{eq:trace_gamma2}.  For $Z_k\equiv 1$ and by taking the
regulator-dependent constant $\ell_1^{(\mathrm F)}$ from
Eq.~\eqref{eq:threshold-constant} into account, we end up with the flows of
the dimensionless couplings $\tilde g$ and $g$
(Eq.~\eqref{eq:dimensionless_couplings}) as displayed in
Eqs.~\eqref{eq:beta-function_tilde-g} and \eqref{eq:beta-function_g}.\\


\end{document}